\documentclass[aps, twocolumn]{revtex4-1}
\usepackage{graphicx}
\usepackage{amsmath}
\usepackage{units}
\usepackage{hyperref}
\graphicspath{ {img/} }

\usepackage{xcolor}

\usepackage{soul}

\usepackage{subcaption}
\usepackage{ragged2e}
\DeclareCaptionJustification{justified}{\justifying}
\DeclareCaptionSubType[Alph]{figure}
\begin{document}
\title{
A novel health risk model based on intraday physical activity time series collected by smartphones}



\author{Evgeny Getmantsev$^1$, Boris Zhurov$^1$, Timothy V. Pyrkov$^{1}$, and Peter O. Fedichev$^{1,2}$}
\address{
$^{1}$Gero LLC, 105064, Nizhny Susalny per. 5/4, Moscow, Russia\\
$^{2}$Moscow Institute of Physics and Technology, 141700,
Institutskii per. 9, Dolgoprudny, Moscow Region, Russia}
\thanks{Correspondence and requestsfor materials should be addressed to P.O.F (email: pf@gero.com)}

\begin{abstract}
We compiled a demo application and collected a motion database of more than 10,000 smartphone users to produce a health risk model trained on physical activity streams. We turned to adversarial domain adaptation and employed the UK Biobank dataset of motion data, augmented by a rich set of clinical information as the source domain to train the model using a deep residual convolutional neuron network (ResNet). The model risk score is a biomarker of ageing, since it was predictive of lifespan and healthspan (as defined by the onset of specified diseases), and was elevated in groups associated with life-shortening lifestyles, such as smoking. We ascertained the target domain performance in a smaller cohort of the mobile application that included users who were willing to share answers to a short questionnaire related to their disease and smoking status. We thus conclude that the proposed pipeline combining deep convolutional and Domain Adversarial neuron networks (DANN) is a powerful tool for disease risk and lifestyle-associated hazard assessment from mobile motion sensors that are transferable across devices and populations.
\end{abstract}
\maketitle

\section*{INTRODUCTION}

Recent progress has been made in identifying and characterizing biomarkers of aging \cite{hannum,horvath,levine2013bioage} 
from a variety of biological signals. Understanding their relation to risks of chronic disease(s) \cite{zhang2016epigenome,pmid25969563,pmid25678027,horvath2014obesity} and lifespan \cite{liu2018phenotypic} creates a new methodology for the performance of large-scale aging studies, clinical trials of anti-aging therapies, health risks assessments (HRA), and informed lifestyle interventions.  Earlier \cite{pyrkov2017quantitative}, we proposed using the massively available intraday physical activity tracks recorded by wearable devices to train proportional hazards models of mortality and morbidity from the National Health and Nutrition Examination Survey (NHANES) activity records. We demonstrated the possibility of quantifying health risks, including the prospective incidence of chronic age-related diseases and death, in the UK Biobank (UKB) cohort. Convolutional Neuron Networks (CNNs) are powerful machine-learning tools that can be used to generate far superior risk models by capturing intricate dependencies among inputs \cite{pyrkov2018extracting}, often, however, at a price of increased sensitivity to small variations in the data. Real world applications of such models could be challenging due to population differences \cite{althoff2017large}, sensor hardware and/or study protocol variations. 
The value of such HRA tools would, therefore, depend on the availability of structured data and on whether the risk models could be effectively trained in a form preserving performance across different signal sources.

Wearable motion sensor technology has reached broad adoption only very recently and hence there are few physical activity records available for longer than $3-5$ years.  The largest dataset available to us, the UK Biobank (UKB), provides nearly $100$K one-week long time series of motion data and a vast registry of well-structured clinical information. Yet the death register in the data is essentially empty due to the limited follow-up and hence is insufficient to train any reasonable all-cause mortality risk model (there are little more than $300$ death events in the relevant cohort!). Besides age and sex, even less information can be found regarding the clinical history or survival of typical mobile tech users.
A popular option involving training a biological age model that predicts chronological age fails to produce a sound health risk model \cite{pyrkov2018extracting}.
Instead, based on recently observed \cite{pyrkov2017quantitative} high concordance between the mortality and morbidity predictors we propose to train health risk models using the far more abundant labels related to the incidence of chronic diseases rather than death.
To achieve generalization of the prediction across the domains represented by motion data collected by different devices and under various experimental conditions, we demonstrate a domain adversarial neuron network (DANN) architecture \cite{ganin2016domain} trained on a large, high-quality structured dataset of motion and clinical information records such as the UKB and, simultaneously on a sufficiently large unlabeled set of activity records representing a cohort of poorly characterized mobile phone users (source and target domains, respectively, in the domain adaptation theory terminology).

We built a demo application for iPhone users, obtained consent, and gathered motion data from nearly $11$K individuals that provided access to their physical activity tracks along with their age and sex. We performed an adversarial domain adaptation of the health risk (morbidity) model trained on the UKB dataset and the motion data from our mobile phone users without any appreciable degradation of the predictor performance on the source domain.
More specifically, we assembled a relatively deep ResNet (see \cite{rajpurkar2017cardiologist} for a recent example of a similar architecture reaching top performance in ECG classification) trained to predict the incidence of at least one of the chronic age-related disease (such as diabetes, heart failure, etc, see Material and methods for the exact definition). The prediction of the network has the form of the log-odds ratio characterizing the probability of being diagnosed with a disease and was used as the model risk score. The risk score is associated with the age of onset of the first chronic disease (the end of the chronic disease-free period or healthspan) and death. For healthy individuals the model correctly assigns elevated risks (and therefore reduced expected healthspan and lifespan) in groups associated with hazardous lifestyles, such as smokers compared with those who quit smoking or never smoked in the respective age- and sex-matched cohorts, c.f. \cite{pyrkov2017quantitative}.
We observed, that the morbidity model risk score negatively correlates with the total activity, albeit the dependence is highly non-trivial:  higher activity levels are expectantly associated with lower health risks, yet the health risk estimate depends on the age of an individual after adjusting for (the negative log of) average activity. In the realm of low activities, we observed distinct high and low-risk groups irrespective of age.
Finally, we ascertained the domain adaptation pipeline performance by validating the prediction using the labels from approximately $1000$ application users who were willing to answer questions related to their health and smoking status.

\section*{RESULTS}

\subsection*{DANN UKB-to-mobile phone users}

\begin{figure*}[htbp]
	\begin{subfigure}{0.38\textwidth}
		\includegraphics[width=\textwidth]{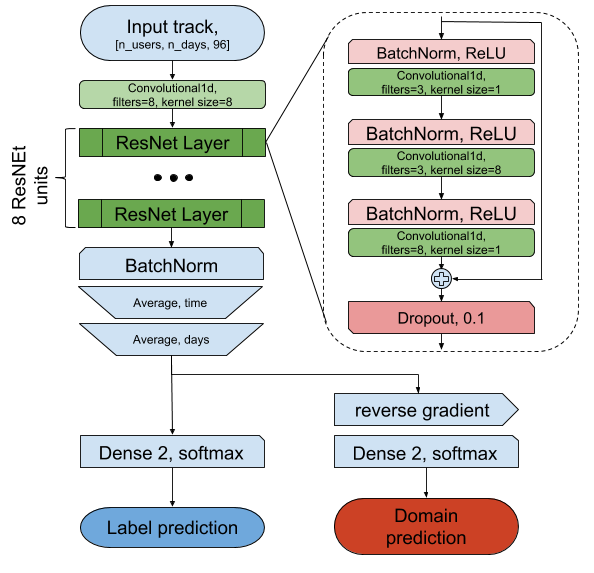}
		\caption{}\label{fig:CNN_architecture}
	\end{subfigure}
    \begin{subfigure}{0.58\textwidth}
    	\includegraphics[width=\textwidth]{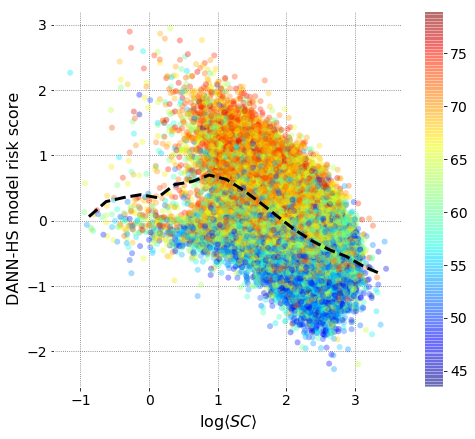}
		\caption{}\label{fig:UKBHZvsLogAct}
    \end{subfigure}
\caption{  
\textbf{\subref{fig:CNN_architecture}.} The domain adversarial neuron network (DANN) includes a deep ResNet classifier trained to predict the disease status of an individual from one-week-long physical activity tracks (left) and the domain label (right).\\
\textbf{\subref{fig:UKBHZvsLogAct}.}
The dependence of the DANN-HS model risk score (the health risk estimate, see the explanation in the text) on the (negative logarithm of) the average activity in the form of the number of steps per minute (step counts, SC) in the UKB. Although there is a profound correlation between the risk and total activity, there is also an apparent correlation between the risk and age at any given activity level. In the low activity region, corresponding to sedentary lifestyle, there are people of low risk irrespective of their age.
}
\end{figure*}

To maximize the practical impact of our research, we gathered the data with mobile phones. As a proof of concept demonstration, we focused on the Apple iOS platform and collected the motion data and the basic demographic information with a freely available application. The focus on a single platform helps by offering a reasonably standardized hardware and software environment. Once the consent is given by a user, the application gains access to a fairly long data stream (on average, every user provided $211$ days of data, $25\%$ of users have at least $21$ non-overlapping non-empty $7$-day segments and $50\%$ have at least $9$). In a recent version of the application we also asked users about their smoking and disease status. Overall, we were able to collect tracks from $10998$ users, from whom $9218$ provided data of sufficient quality and were in the age range $20-85$. Of these users, the smoking status variable was available for $670$, morbidity for $715$ and both labels for $663$ individuals (see Materials and Methods).

Wearable motion sensors vary in their hardware and the signal processing software and therefore the distribution of physical activity recorded by different devices may arise from similar but different distributions. 
Domain-Adversarial Neural Network (DANN) architecture~\cite{ganin2016domain} is directly inspired by domain adaptation theory  \cite{ben2010theory}, can be used to train a supervised model on the labeled (source) domain and, at the same time, construct features that cannot discriminate between the source and target domains, see Fig.~\ref{fig:CNN_architecture}. 
In this work, we used a large and open dataset, the UK Biobank (UKB), which provides nearly $100$K one-week long physical activity records (a tri-axis accelerometer at $100$Hz sampling frequency), augmented by a vast registry of well-structured clinical information, as the source domain for DANN. 

The data from wearable sensors, however, were only recently added to the UKB and the short follow-up precludes training of any reasonable health risk model in the form of a mortality prediction (there are a little more than $300$ death events). In \cite{pyrkov2017quantitative} we observed that the log-hazard ratio of a morbidity model could be viewed as a dynamic frailty index that is closely associated with biological age. 
Therefore, it should be possible to produce a biological age version in the form of a log-proportional hazard model trained to predict the age at the onset of major age-related diseases. 
Since the incidence of the most prevalent chronic diseases in the UKB increase exponentially with age, it would be natural to employ a parametric risk Cox-Gompertz proportional hazard model \cite{zenin2018identification, pyrkov2018extracting}. 
If the number of events in a Cox-Gompertz proportional hazard model is small, the log-linear risk predictor coincides approximately with the log-odds ratio of a logistic regression classifier trained to predict the incidence of the same events \cite{green1983comparison, abbott1985logistic}. 

Following \cite{zenin2018identification,pyrkov2017quantitative}, we classified UKB participants as disease-free if they were not diagnosed as having any of the following health conditions: cardiovascular (angina pectoris, coronary heart disease, heart attack, heart failure, stroke), diabetes, hypertension, arthritis, emphysema or cancers at the time that physical activity was recorded. The list of diseases in the UKB was collected from individual ICD10 labels and was refined to overlap with that of the NHANES (see details in Materials and Methods). The total fraction of UKB participants diagnosed with at least one of the relevant diseases is much larger than the number of mortalities ($51035$ vs. $302$ individuals in total, or $53.5$\% and $0.32$\% of the study participants, respectively). Therefore, the combination of the two measures: using healthspan instead of lifespan as a target phenotype and switching to a binary disease-label classifier instead of a proportional hazards model, should help to at least improve the statistical power of the resulting model.  The approach should be especially helpful in situations where it is hard or impossible to obtain the exact date of the diagnoses and/or the diagnoses follow-up data.

The DANN employed in this work includes a deep convolutional neural network (CNN), see Fig. \ref{fig:CNN_architecture} that was trained in batches consisting of equal amounts of tracks from the source and target domains. To make an optimization of such a deep network feasible, we followed \cite{rajpurkar2017cardiologist} and turned to a ResNet architecture using the full preactivation variant from \cite{he2016identity}. The shortcut connections between the layers help training by letting information propagate well across the depth of the network. The CNN consists of $8$ residual units with $3$ convolutional and one drop-out layer per unit, see Fig.~\ref{fig:CNN_architecture}. Note that we replaced the batch normalization layers used in \cite{rajpurkar2017cardiologist} with the simple per-component linear transformations, see Materials and Methods for details. 
The higher level ResNet representation of the input signal is used simultaneously for the morbidity label prediction on the source domain (UKB) and for the domain label differentiation (with the weighted reverse gradient) as in \cite{ganin2016domain}, respectively. 

\begin{figure}[htbp]
\includegraphics[width=0.99\columnwidth]{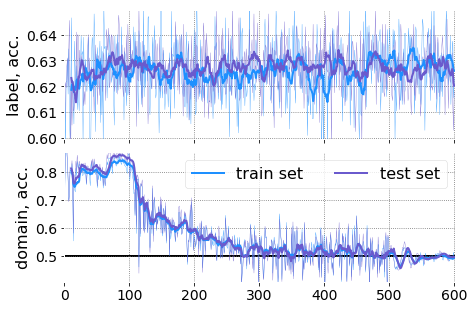}
\caption{Learning curves for DANN. Accuracy for label and domain prediction for both train and test subsets are shown. Thick lines are running averages for 10 epochs, thin lines are raw values. Domain prediction accuracy $0.5$, corresponds to the inability to distinguish domains is highlighted. Note that for each epoch accuracy was calculated using 2000 randomly selected samples.}
\label{fig:LCurve}
\end{figure}

The network was trained from scratch with TensorFlow~\cite{tensorflow2015-whitepaper}. We used default initialization for the weights of the layers and Adam optimizer~\cite{kingma2014adam} with default parameters. We let the network train on the source domain for $100$ epochs first, then gradually turned on the inverse gradient from the domain predictor over $300$ epochs, thus forcing the domain inseparability, followed by another $200$ epochs of training with decreasing learning rate to reach the stable model. The network learns the data representation indistinguishable between the source and the target domains without any apparent degradation of the source domain performance, see Fig.~\ref{fig:LCurve} for the learning curves. 
We employed the best model according to its performance on the training set during the last $25$ epochs of optimization. At this stage, the network had reached a steady state without significant differences in accuracy between training and test sets.
The reported architecture was the result of an extensive grid search involving variation in the number of the ResNet units, as well as the number and parameters of convolutional layers within ResNet units, dropouts, and non-linearity types.

\subsection{Source domain model performance}
\begin{table*}[htbp]
\caption{
Validation of the health risk model scores in the prospective morbidity and mortality models. The significance tests were performed using a series of increasingly sophisticated Cox-Gompertz proportional hazard models including Basic (age, sex), Advanced (age, sex, $\log\langle SC\rangle$), and Extended (age, sex, BMI, smoking, $\log\langle SC\rangle$). Here SC (step count) is the number of steps per minute and $\log\langle SC\rangle$ represents the average total activity of an individual. $^*$Smoking status and body mass index (BMI) are not available for all users, so the numbers are about $5\%$ smaller for the Extended model.
}
\begin{tabular}{lrrr|rr|rr|rr}
\hline
DNN & Dataset & size$^*$ & events$^*$ & HR & p-val & HR & p-val & HR & p-val \\
\hline
\multicolumn{4}{r}{Baseline morbidity hazard:} & \multicolumn{2}{|l}{Basic} & \multicolumn{2}{|l}{Advanced} & \multicolumn{2}{|l}{Extended} \\
\hline
UKB-HS & UKB(train) & 36524 & 987 & 1.399 & 7e-25 & 1.345 & 8e-20 & 1.274 & 2e-13 \\
UKB-HS & UKB(test) & 9164 & 262 & 1.285 & 9e-05 & 1.188 & 6e-03 & 1.131 & 5e-02 \\
DANN-HS & UKB(train) & 36524 & 987 & 1.385 & 1e-27 & 1.347 & 1e-22 & 1.283 & 6e-16 \\
DANN-HS & UKB(test) & 9164 & 262 & 1.342 & 5e-07 & 1.257 & 1e-04 & 1.196 & 3e-03 \\
\hline
\multicolumn{4}{r}{Baseline mortality hazard:} & \multicolumn{2}{|l}{Basic} & \multicolumn{2}{|l}{Advanced} & \multicolumn{2}{|l}{Extended} \\
\hline
UKB-HS & UKB(train) & 76379 & 238 & 2.060 & 1e-29 & 1.498 & 1e-11 & 1.454 & 1e-09 \\
UKB-HS & UKB(test) & 19095 & 64 & 2.103 & 1e-09 & 1.719 & 7e-06 & 1.705 & 1e-05 \\
DANN-HS & UKB(train) & 76379 & 238 & 1.838 & 2e-31 & 1.373 & 7e-10 & 1.335 & 3e-08 \\
DANN-HS & UKB(test) & 19095 & 64 & 1.839 & 7e-09 & 1.541 & 6e-05 & 1.523 & 1e-04 \\
DANN-HS & NHANES(full) & 4294 & 240 & 1.371 & 4e-13 & 1.083 & 6e-02 & 1.101 & 3e-02 \\
DANN-HS & NHANES(40+) & 2391 & 229 & 1.428 & 1e-12 & 1.101 & 5e-02 & 1.117 & 2e-02 \\
\end{tabular}
\label{table:PValAll}
\end{table*}

Following \cite{pyrkov2017quantitative}, we proposed the log-odds ratio of the ResNet classifier after the domain adaptation (hereinafter referred to as DANN-HS model) as a natural definition of the risk variable, similar in properties to biological age. Fig.~\ref{fig:UKBHZvsLogAct} is a plot of the risk score as the function of the (logarithm of) total activity (each dot represents one UKB participant and the colors are assigned by the age). 
First, we observe, that the risk score is positively associated with age (Pearson's correlation coefficient, $R=0.42$), and is negatively associated with (logarithm of) total (or average) activity, ($R=-0.54$).
Although there is a profound correlation between the risk and total activity (see the risk score reducing from left to right on the figure irrespective of the age), there is also a significant correlation between the risk and age at any given activity level. What is most remarkable, however, is that there are people of low risk and low activity irrespective of their age. The conclusion is one of the central results of our study: the model appears to differentiate the sub-populations of significantly and, in the lowest activity range, qualitatively different risks at the same level of total physical activity.

More formally, we tested the significance of the excess of the risk, or the model risk acceleration, understood as the difference between the risk estimate and its mean value in a age-and sex-matched cohort \cite{pyrkov2017quantitative, pyrkov2018extracting}, in prospective mortality and morbidity models. We used a log-likelihood ratio to characterize the DANN-HS model risk predictor in proportional parametric (Cox-Gompertz) hazard models employing the age, sex and a number of additional covariates (see DANN-HS entries in Table \ref{table:PValAll}). First, we observed that there is no performance degradation of the model in the source UKB dataset. To see this we provide the performance metrics for the additional UKB-HS model, that used the same ResNet architecture trained to predict the disease label on the source domain UKB entries only. In fact, the DANN-HS model has a marginally better performance on the UKB.

The reference variants of the proportional hazard models use age and sex as the model covariates for mortality and morbidity prediction. The proportional hazard regression coefficients are $0.10$ and $0.059$ per year for age corresponding to reasonable $6.9$ and $12$ years of the mortality and morbidity rate doubling times, close to the accepted value of $8$ years for the mortality rate doubling time from the Gompertz mortality law \cite{gompertz1825nature, makeham1860law}. Under the Gompertz law, the models yield $98$ and $70$ years of average lifespan and healthspan. While the healthspan estimate looks good, the lifespan is inflated, which is apparently the consequence of the limited number of deaths recorded in the dataset. The models predict $6.6$ and $3.6$ years of lifespan and healthspan difference, respectively, between males and females.

We added the DANN-HS risk score to the reference hazard models and saw a very significant association with both the remaining healthspan and lifespan (see ``Basic'' model table entries). 
The associations of the DANN-HS risk score remains significant after further adjustment by the negative logarithm of the average activity (a convenient measure of activity strongly associated with frailty and biological age \cite{pyrkov2017quantitative}; see the corresponding ``Advanced'' model table entries). More importantly, the association of DANN-HS risk score with the remaining health- and lifespan does not lose its significance after further adjustments by smoking status and body mass index (BMI), another two factors related to life-shortening lifestyles. The Cox regression coefficients can be used to rescale the DANN-HS risk score into the remaining healthspan in years (under the Gompertz law assumption).

\begin{table*}[htbp]
\caption{Shift of expected healthspan due to smoking in healthy individuals. Locomotor log-hazard was detrended by subtracting the mean log-hazard in the age-sex matched cohort. P-values from Mann–Whitney U test for difference between 'N' (never smoke), 'Q' (quit smoking) and 'C' (current smokers) cohorts are shown.}
\begin{tabular}{l|r|r|r|r|r|r|r|r}
\multicolumn{2}{l}{} & \multicolumn{3}{|l}{Group size} & \multicolumn{3}{|l}{\begin{tabular}{@{}c@{}}diff. between groups \\ $\Delta(HS)$, years (p-value)\end{tabular}} \\
DNN & Dataset & N & Q & C & N vs Q & N vs C & Q vs C \\
\hline
UKB-HS & UKB(train) & 22382 & 11589 & 2553 & -0.23 (5e-03) & -2.17 (3e-36) & -1.93 (4e-27) \\
UKB-HS & UKB(test) & 5596 & 2910 & 658 & -0.31 (7e-02) & -2.10 (6e-10) & -1.79 (2e-07) \\
DANN-HS & DANN(train) & 22382 & 11589 & 2553 & -0.30 (1e-06) & -1.96 (9e-55) & -1.67 (2e-36) \\
DANN-HS & DANN(test) & 5596 & 2910 & 658 & -0.27 (4e-02) & -1.84 (7e-13) & -1.57 (3e-09) \\
DANN-HS & NHANES & 1157 & 385 & 448 & -0.29 (2e-01) & -1.24 (4e-07) & -0.95 (3e-04) \\
DANN-HS & HK(phone) & 338 & 115 & 79 & -0.98 (8e-02) & -2.49 (2e-04) & -1.51 (2e-02) \\
\end{tabular}
\label{table:SmokeSplits}
\end{table*}

The identification of log-odd ratio from the neural network trained to predict morbidity event as the general health-related risk score is further supported by comparing the healthy (chronic diseases-free) individuals. The model correctly assigns the elevated risks (and hence reduced life- and healthspan) for current smokers in age- and sex-adjusted cohorts both in the train and test sets (see respective entries in Table~\ref{table:SmokeSplits}). The effect appears to be almost entirely reversed in the groups of the individuals who quit smoking, as expected c.f. \cite{pyrkov2017quantitative}. Once again, DANN-HS actually perform slightly better than the simple UKB-HS model (see Discussion for possible reasons behind such behavior).

Qualitatively, the DANN model can be applied to an independent dataset without additional adaptation. In the NHANES dataset, we see a significant association of the DANN model risk score with mortality (see DANN-HS NHANES (full and 40+) entries in Table~\ref{table:PValAll}) and with smoking status in age- and sex-matched cohorts of healthy individuals (see Table~\ref{table:SmokeSplits}).The effect size in the lifespan determination and in association with smoking is significantly smaller than that in the UKB. Nevertheless, based on the results presented in this section, we conclude that DANN-HS model provides a reasonable health risk estimate concordant with the remaining health- and lifespan and is informative of hazardous lifestyles.

\subsection{Target domain performance}

The association of the DANN-HS risk score with life-shortening lifestyles or diseases suggests a simple way to conduct the model validation on the target domain. We asked a limited number of our mobile application users two questions regarding their chronic diseases and smoking habits ($733$ and $688$ responses, respectively). Among healthy individuals, the DANN-HS model correctly assigns higher risks to current smokers compared to there age- and sex-matched peers in concordance with the results seen in the UKB and NHANES (see ``HK(phone)'' row in Table~\ref{table:SmokeSplits}). We also observed higher risks for the people diagnosed with chronic diseases individuals compared to there age- and sex-matched (self-reported) disease-free peers (respectively, $177$ and $572$ individuals with known sex and known to be between $20$ and $80$ years old, Mann–Whitney U test p-value $0.05$).

\begin{figure}[htbp]
\includegraphics[width=0.45\textwidth]{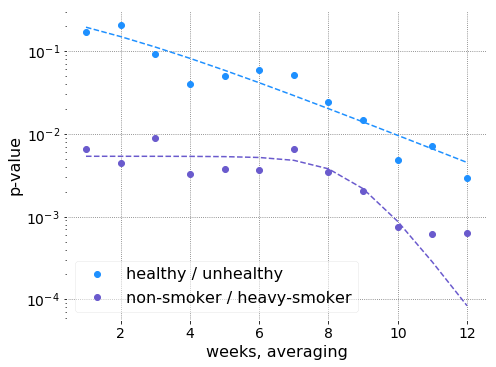}
\caption{
Significance of age- and sex-adjusted track averaged DANN-HS model risk score stratification between groups of heavy smokers vs. never smokers and healthy vs. users with a disease as a function of the averaging time. The dotted lines are guides for eyes.
}\label{fig:pval-averaging}
\end{figure}

Technically, DANN-HS model is trained using one-week long physical activity records and therefore could be reported once a week. Alternatively, one can produce a health risk assessment every following week and report the average obtained over longer periods in the form of a sliding average. While it would be impossible to assess the quality improvement as a result of such a procedure in the UKB, the longitudinal character and the accessibility of HealthKit motion data can be used to understand the time scales involved in the model health score and hence the biological age or frailty index variation. In Fig.~\ref{fig:pval-averaging} we plotted the negative log of Mann-Whitney U-test p-values between the groups representing heavy smokers (those who smoke every day) vs. those who never smoked and the disease-free vs. people diagnosed with chronic diseases, as a function of the length of the track used for the DANN-HS score evaluation. 
The significance of the separation improves sharply once the track exceeds approximately $10$ weeks (note the $\log$-scale along the vertical axis).

Finally, we observed that the results of the inter-group separation between tracks of smokers and non-smokers comparing users) do not depend on the order of the averaging: the effect and significance of the healthspan estimate turned out to be the same between the log-hazard ratio predictors averaged along all the short tracks of a user, or between the age- and sex- matched averages of all the available short tracks.

\section*{DISCUSSION}


We report a successful demonstration of training and validation of a health risk model based on motion sensor data from the most widely available source of biological information, the mobile phone. We proposed to overcome the lack of medically labeled data in the following steps. First, we used a large UKB dataset combining a slice of motion data and a well-structured set of clinical variables as a training set for the model. Second, we used the association of the biological age, understood as the risk factor behind the incidence of chronic diseases and mortality, with frailty and suggested that we train a binary classifier predicting the incidence of age-related diseases in an individual, instead of a more commonly applied proportional hazards mortality or morbidity model. Finally, we employed adversarial domain adaptation technology in combination with deep residual convolutional networks to automate the extraction of relevant features from the time series and the building of a non-trivial risk model that performs well across different sources of motion data. We developed a simple proof-of-concept application for mobile phone users, gathered enough tracks for the model adaptation and collected enough health and lifestyle information from a small cohort of users to validate the model and hence the whole domain adaptation pipeline.

Remaining lifespan or all-cause mortality prediction is the quintessential measure of the overall organism health, closely related to the traditional frailty index and emerging concept of biological age \cite{pyrkov2017quantitative}.
Switching to a highly concordant measure, the healthspan, and the associated risks of morbidity in the form of incidence of chronic diseases of age offers multiple advantages for the training of health risk models. 
First of all, the wearable technology is still relatively young and hence the disease status is a far more abundant label in any realistic dataset (there are more than $50\%$ of UKB participants diagnosed with at least one disease of age), whereas the number of death events is very low due to short follow-up time. 
The incidence of chronic diseases associated with aging increases exponentially with age at a rate compatible with the mortality rate doubling rate of Gompertz law for most diseases \cite{zenin2018identification}. 
If the prevalence of these diseases is small,  parametric Cox-Gompertz proportional hazards model are formally equivalent to a logistic regression classifier trained to predict the diseases status of an individual. That approach is adopted here and offers another important advantage: one does not need to know the exact age at the onset of the diseases, which is almost always not known exactly.
As more data become available, including the motion sensor streams from older individuals, the prevalence of the diseases would increase and it could be reasonable to switch to full proportional hazards models. Fortunately, the exponential character of the diseases incidence acceleration with age lends a simple maximum likelihood for the Cox-Gompertz risk model allowing for an efficient integration into deep learning architectures, see \cite{pyrkov2018extracting}. 

The results presented here should, in principle, depend on the exact definition of healthspan or, more specifically, on the list of diseases we use to define an individual as disease-free and use such a label to train the morbidity predictor. It should be noted, however, that there is almost never an extended time period characterized by a single disease. 
A first morbidity is quickly followed by another within a few years \cite{st2015risk} and therefore the exact definition of the healthspan termination event should not be particularly important. We expect that the definition will be refined in the future and would lead to achieving incremental model improvements.

By modern standards, the best morbidity model achieved only a fair accuracy of the label assignment on the source data set (UKB). However, the morbidity model is by its very nature the risk model \cite{pyrkov2017quantitative}. The model cannot identify the incidence of the specific diseases (at least in its current form) and therefore estimates the general health score of the person. The very nature of such generic organism-wide health score leads to a fundamental impossibility to separate (on the basis of such score alone) healthy people with life-shortening lifestyles from individuals with a disease.  

The non-trivial dependence of the morbidity model risk score on the total activity and chronological age, see Fig.~\ref{fig:UKBHZvsLogAct}, is one of the central results of this study. The DANN-HS model produces the expected association of the risk predictor and the total activity level (more precisely, the log-odds or hazard ratio and the negative logarithm of total activity \cite{pyrkov2017quantitative}). In the low activity range, however, the model outputs a risk estimate different in groups representing frail and chronic disease-free individuals irrespective of their activity levels. A more careful inspection shows that the dependence of the log-risk ratio on age remains significant at any given level of the total activity. Moreover, it appears harder for an individual to support high levels of physical activity at advanced ages.

The dependence of the health risk on the total physical activity is a genuine challenge, particularly if models are developed using variables representing physical activity itself. The difficulty, however, is not limited to biomarkers derived from the motion data alone. For example, traditional physiological state variables, such as blood cell counts, would also depend on the overall level of activity and sedentary lifestyle through basic metabolic rate and associated hemoglobin levels, a signature of oxygen consumption \cite{willis2018association}. Resting heart rate is also associated with the total activity \cite{jensen2013elevated}. 
Physical activity levels in developed countries are on average lower and yet those countries maintain the highest life expectancy \cite{althoff2017large}. 
In the future, the proper disentanglement of the health risk and the total activity will be achieved as motion data accumulates from individuals from vastly different lifestyles and is integrated into the training of the models.
The longitudinal character of the mobile phone motion dataset allowed us to glance into the dynamic properties of the biological age and morbidity risk estimates, see Fig. \ref{fig:pval-averaging}. For practical reasons, we produced predictions using every available week of data, both for the biobank and the mobile phone users. For the latter, however, we could generate as many risk estimations as the number of weeks of the motion data stored in HealthKit. We compared the risk scores averaged over short users trajectories vs. the full-length estimates and observed, that the score begins averaging out after about $8-10$ weeks of data. We, therefore, conclude that there are fluctuations in the risk score with relatively high-frequency and short auto-correlation times which can be reduced by averaging over time scales exceeding $2-3$ months. It remains to be seen, however, especially in view of future anti-aging clinical trials design, what is exactly the response time of the dynamic variable most intimately associated with the biological age and frailty. 





In this study, we applied a popular deep residual CNN architecture, a ResNet \cite{he2016deep}. A similar network has recently used for human-level quality of ECG classification in \cite{rajpurkar2017cardiologist}, which served as an inspiration for this work. The approach provides a sufficient level of model sophistication and non-linearity to produce a rich set biological state representations associated with the diseases and mortality risks. The architecture could be naturally used in a DANN along with the domain classifier trained with a reverse gradient to produce a narrower set of features suitable for the domain adaptation \cite{ganin2016domain} of the morbidity model from the UKB to a different source of the signal; the mobile phone users' generated activity dataset. We hope that the same architecture could be used to train more sophisticated and accurate models once more motion data is available with only incremental modifications. 

To produce a better alignment between the source and the target datasets prior to DANN procedures, we dropped some user from source and target domains to achieve the same age distribution among remaining users. We also transformed the UKB data to $1 min^{-1}$ sampling frequency with the help of a simple ``step counter'' to match the data available in the mobile phone activity time series (see Materials and Methods). We further down-sampled the data in both domains to account for sparsity of the data actually available in the mobile phone, which led to a further loss of information from the source domain. Many continuously wearable devices, such as fitness trackers, provide tracks at $1 min^{-1}$ sampling frequency, hence we expect that the same technology as described here may be employed with such higher quality motion data to yield considerably more precise risk models. Further performance gains could be attained by building even higher sampling frequency devices and applications or, additionally, by augmenting the signal by the continuous heart rate monitoring.  

Our mobile application users showed a remarkable degree of engagement and willingness to share their motion data along with their basic demographic and health/lifestyle information. While it is still hard to assemble a sufficiently large database with quality health details, the mobile app allowed us to collect just enough self-reports to validate the DANN risk model. At the moment, our best model is deployed on Apple iOS in view of the apparent homogeneity of the software and hardware among its users. This is not, however, a hard limit: in the future we expect DANN to help to bring more hardware and software platforms for health risk assessment.

The relationship between the log-health risk ratio and the biomarker of age suggested in \cite{pyrkov2017quantitative} and extended here, offering a novel exciting possibility of using massively available motion data from wearable and mobile phone devices for monitoring associations of lifestyles and healthspan. The digital biomarker can find immediate applications in life insurance. It could also be used to gauge the effect of lifestyle modifications through nutrition and exercise, as of now, and in response to anti-aging diet plans, supplements and therapies, in the future. Future observational studies in a sufficiently large cohort labeled by potentially life- and healthspan extending interventions could eventually lead to another level of AI-driven fully automatic recommendation systems tailoring anti-aging treatments and lifestyle interventions depending on the individual life-history variables produced by wearable devices.

\section*{Materials and methods}

\subsection{HealthKit dataset}\label{MM_HK}
The smartphone-base dataset was collected in $2016-2018$ using publicly available application [https://apple.co/2OHaOAQ]. We used HealthKit API [https://developer.apple.com/healthkit/] to collect physical activity as measured by users' iPhones (this dataset is labeled ``HK(phone)'' or simply ``HK'' in this paper). Users were asked to provide there age and sex ($4.5\%$ refused to do so). Starting from September $2018$ users ($972$ as of $2018-10-22$) were asked questions about their smoking and morbidity status (see exact formulations below).

Both UKB and NHANES (two other datasets used in this study) had 7-day long tracks so we chose to split HK tracks into 7-day slices as well. HK users were carrying smartphones casually, and as a result, some days contained too few data points to be usable. Also internally iPhone stores step counts per non-equal intervals ranging from few seconds to more than an hour ($15\%-75\%$ range is from $5 s$ to $9.8 min$ with $5.28 min$ on average) and the steps-per-minute track was an interpolation available through HealthKit API. This sometimes resulted in days with constant non-zero step counts over long periods. We decided to combat these issues by filtering out ``bad'' days: days with less than $150$ minutes with non-zero step count, days with less than $45$ significant ($>1$) changes and days with less than $10$ different step count values were dropped. On average, the HK tracks we collected were $211$ days long (from the first day with some data to the last), of which $158$ days contained some data and $106$ days which passed the filter. The filter parameters were obtained as a compromise between hard filtering leaving too few days and soft filtering resulting in too few data within slices (filter parameters were obtained during preliminary studies not presented here). With filtering we got only $9623$ users with $98402$ non-overlapping continuous 7-day slices in total. We decided to relax the filtering by allowing 7 ``good'' days to be selected from up to 14 days. This gave us $10998$ users with $164633$ non-overlapping 7-day slices, including $9218$ users known to be between $20$ and $85$ y.o. with smoking status known for $688$ and morbidity status for $733$ users. The average length of the continuous slice from which the 7 day slice was extracted was $8.3$ days.

\subsection{NHANES dataset}\label{MM_NHANES}
The National Health and Nutrition Examination Survey (NHANES) data were downloaded from {[}\url{www.cdc.gov/nchs/nhanes/index.htm}{]}. The locomotor activity was available in NHANES in the form of ``activity counts'' and step counts (both per minute). We chose the later for compatibility with HK. The step counts were available for $6926$ participants from $2005-2006$ cohort. The data were collected by ActiGraph AM-7164 single-axis piezoelectric accelerometer worn on the hip for 7 consecutive days. We excluded participants with abnormally low physical activity (those with less than $250$ minutes with at least $4$ steps) and participants aged $85$ (since the \href{https://wwwn.cdc.gov/Nchs/Nhanes/2005-2006/DEMO_D.htm#RIDAGEYR}{NHANES age data field} is top coded at 85 years of age and we desired precise age information for our study) leaving $6635$ participants (including $3863$ aged $20+$ and $2391$ ages $40+$). The mortality data for NHANES participants was obtained from the National Center for Health Statistics \href{https://www.cdc.gov/nchs/data-linkage/mortality-public.htm}{public resources} and was available for $4294$ participants, including $3862$ aged $20+$ and $2391$ ages $40+$. The mortality data follow-up time was between $0.2$ and $7$ years with $5.9$ years on average.

\subsection{UKB dataset}\label{MM_UKB}

We accessed data from UK Biobank (UKB) under the \href{http://www.ukbiobank.ac.uk/2015/06/dr-peter-fedichev-quantum-pharmaceuticals-moscow/}{approved research project 21988 (formerly 9086)}. At the time the present study was conducted (2018), locomotor activity data were collected in $2013-2015$ for $103711$ UKB participants. In addition to questionary data and measurements made during surveys, UKB contained records from the death registry and hospital records with ICD-10 coded diagnoses collected up until February 2016. The mortality and morbidity data follow-up times for the locomotor activity were between $0.2$ and $2.8$ years, averaging at $1.3$ years.

The physical activity was measured using Axivity AX3 tri-axial accelerometers worn on the wrist for 7 consecutive days. Raw UKB tracks comprised 3D acceleration sampled at $100Hz$. This is very different from steps per minute available in HK. Since data collected from a casually carried smartphone and continuously worn device would never match exactly and domain adaptation would be required anyway, the high quality transformation of the high-frequency acceleration tracks into steps per minute was not needed. We estimated the number of steps as the number of well-separated (at least $250 ms$ apart), large ($\Delta a\geq0.2g$ where $a$ is the acceleration magnitude) and narrow (at most $500 ms$) peaks (see Appendix XXX for more technical description). These ``step counts'' were by now means exact, but they were good enough for DANN.

We managed to transform the raw 3D acceleration into step-per-minutes for $102971$ participants out of $103711$ with locomotor activity data (we dropped tracks with more than $10$ ECC errors inside as well as tracks with discontinuities within internal timestamps; also, some of the raw tracks were actually empty). We checked data in the registry for consistency between the birth date, date of survey, age at survey, date of death (if available) and the date of the start of the locomotor activity measurements. Participants without any of these data fields or with any inconsistency among them we dropped from consideration. We also dropped participants unable or unwilling to share information about their diabetes, cancer or smoking status, leaving us with $102245$ participants. We further filtered out participants with less than $7$ days of activity record and with abnormally low physical activity (less then $250$ minutes with at least $4$ steps) leaving as with $95474$ participants used in this study. The age range (at the time of locomotor data collection) was from $43$ to $79$ y.o. with $63$ y.o. on average.

\subsection{Morbidity label definition}

\begin{table*}[htbp]
  \caption{Number of events derived from clinical and interview data for selected heath conditions.}
\begin{tabular}{lrrrrrrrrrr}
\toprule
{} & \multicolumn{2}{l}{UKB (ICD10)} & \multicolumn{2}{l}{UKB (Self report)} & \multicolumn{2}{l}{UKB (Combined)} & \multicolumn{2}{l}{NHANES (Age 40-85)} \\
{} &                   events &     \% & events &     \% & events &     \% &   events    &     \% \\
hypertension       &    13817 &   27.0 &  21667 &   42.3 &  25380 &   49.5 &     1039    &   67.4 \\
arthritis          &    14145 &   27.6 &  15692 &   30.6 &  24321 &   47.5 &     1443    &   54.2 \\
cancer (all kinds) &    12155 &   23.7 &   8444 &   16.5 &  14146 &   27.6 &      273    &   17.7 \\
diabetes           &     3030 &    5.9 &   3420 &    6.7 &   4294 &    8.4 &      342    &   22.2 \\
CHD                &     5047 &    9.8 &   4636 &    9.0 &   7049 &   13.8 &      134    &    8.7 \\
MI (heart attack)  &     1307 &    2.6 &   1898 &    3.7 &   2443 &    4.8 &      130    &    8.4 \\
angina pectoris    &     2856 &    5.6 &   2093 &    4.1 &   3638 &    7.1 &      106    &    6.9 \\
stroke             &      706 &    1.4 &    931 &    1.8 &   1350 &    2.6 &      102    &    6.6 \\
emphysema          &     1029 &    2.0 &    788 &    1.5 &   1606 &    3.1 &       67    &    4.3 \\
CHF                &      640 &    1.2 &     50 &    0.1 &    662 &    1.3 &      111    &    7.2 \\
\end{tabular}
\label{table:Morbidity_label}
\end{table*}

The health risk model was trained on the source domain (UKB training set) to predict morbidity status. We followed \cite{andersen2012health} and selected the top ten morbidities strongly associated with age after the age of $40$. NHANES Data on having a health condition and age when a participant was told they have a health condition is available in questionnaire category ``Medical Conditions'' (MCQ). Data on diabetes and hypertension was retrieved additionally from questionnaire categories ``Diabetes'' (DIQ) and ``Blood Pressure \& Cholesterol'' (BPQ), respectively. 

UK Biobank does not provide aggregated data on these medical conditions, instead it provides self-reported questionnaire data and a detailed data categorized according to ICD10 codes. We aggregated self-reported and ICD10 (block level) data to match that of NHANES, since testing for transferability of the results between populations and datasets was one of the main goals of our study. With that aim we used the following ICD10 codes to cover the health conditions in UK Biobank: hypertension (I10-I15), arthritis (M00-M25), cancer (C00-C99), diabetes (E10-E14), CHD (I20-I25), MI (I21, I22), angina pectoris (I20), stroke (I60-I64), emphysema (J43, J44), CHF (I50). We also used corresponding self-reported data. 

The morbidity status of a participant was then defined as a binary label indicator of having one of these major age-related health conditions. The list of the health conditions included cancer (any kind), cardiovascular conditions (angina pectoris, coronary heart disease, heart attack, heart failure, stroke, hypertension), diabetes, arthritis and emphysema. We did not consider any mental health conditions like dementia because we did not find any significant association of such conditions with age. The number of events of each of the health condition and the corresponding percentage relative to aggregated morbidity status are given for UKB and NHANES in Table \ref{table:Morbidity_label}. Total number of morbidity status detected for UK Biobank and NHANES (age 40-85) participants was 51243 and 1542, respectively (number of users 95605 and 2282).

\subsection{Adversarial domain adaptation}

In the present study we used domain-adversarial neural network (DANN) for learning the morbidity label in UKB dataset and transfering it to our smartphone-based HK datset. We had two important issues to address before selecting an architecture. First, the slices from HK were not continuous, so we chose to process each day separately and then combine single day feature vectors into a full feature vector before final classification. Second, the original iPhone-collected step counts did not have a $1 min$ resolution, so we decided to downsample tracks using $15 min$ averaging, thus representing each day with $96$ values. Our preliminary studies (not provided here) using UKB data indicated, that the downsampling resulted in less then $1\%$ loss in accuracy.

\subsubsection{Network architecture}

The network architecture is outlined in Fig. \ref{fig:CNN_architecture}. We selectected the optimal architecture for this deep learning model empirically, after exploring a range of combinations of various layer depths and sizes. We chose the ``full pre-activation'' variant of ResNet architecture proposed in \cite{he2016identity} with $8$ identical ResNet units, each  including  a batch normalization layer followed by a ReLU (or other non-linear activation) before each convolutional layer with all convolutional layers having linear activation. We tried both simple batch normalization from \cite{ioffe2015batch} and its advanced variants from \cite{ioffe2017batch}. We also attempted using dummy ``batch normalization`` layers, that were effectively linear transformations for all batch normalization layers and per-component liner transformation as the first batch normalization layer in ResNet unit with identity transformation for the rest. Batch normalization is usually beneficial and it was somewhat surprising that the networks with liner transformations instead of ``batch normalization'' had the best performance. Several factor may contributed to this effect: a large difference between source and target domains distributions (especially in the first ResNet units), large dropout rates causing troubles with batch normalization \cite{li2018understanding}. We had to use relatively deep ResNet to achieve a reasonable performance for DANN and yet the inherent regularization of batch normalization layers was insufficient to prevent the overfitting. We decided to use the best architecture identified during grid search and left a more thorough investigation of the batch normalization issues for a subsequent study.

In our architecture the tracks with the shape $N\times96$ where $N=N_{batch}\times N_{days}$ go through the convolution layer of size $8$ with $8$ filters, followed by $8$ ResNet units. Each ResNet unit had $3$ convolution layers of size $1$, $8$, $1$ with $3$, $3$ and $8$ filters. Before each convolution layer there was a per-component (along the filters axis) linear transformation followed by a ReLU activation. During training, $10\%$ dropout was applied after each ResNet unit. After ResNet units the data have shape $N\times96\times8$ and go through batch normalization followed by average pooling producing $N\times8$ daily descriptors. The daily descriptors are then reshaped into $N_{batch}\times N_{days}\times8$ and converted to $N_{batch}\times8$ user descriptors with another average pooling. The user descriptors are fed to dense layer with two neurons followed by softmax layer giving the cost for label prediction. The domain adversarial part of the network had inverse gradient layer with scaling $\lambda$ followed by dense layer with two neurons and softmax giving the cost for domain prediction.
When training DANN, only descriptors for users from UKB (the source domain) were fed to the label prediction part of the network.

For reference, we also trained a ResNet on UKB tracks to see how much accuracy was lost as a result of domain adaptation. The network we used had the same architecture as DANN above, only without domain prediction part.

\subsubsection{DANN training}

In this study we used the morbidity event as the label to be predicted. Aging is the major contribution to the morbidity and a descriptor useful for predicting morbidity will naturally be age-dependent. Since UKB and HK had a very different age distribution, the average descriptor for UKB user would differ from HK one. This posed a problem for DANN, since, by design, it tries to make the descriptor indistinguishable between datasets. We addressed this problem by balancing the age distribution in our datasets before training DANN.
We split age range into $5$ year cohorts from $45$ to $85$ y.o. We dropped all HK users not in $45-85$ age range (these users had no age-matching peers in UKB), excess users from $45-50$ cohort from HK (target domain) and excess users from in $55-85$ cohorts from UKB (source domain), thus obtaining source and target domain subsets with identical distribution per $5$ year age cohorts.

We used the Adam \cite{kingma2014adam} optimizer. For the first $100$ epochs the domain predictor inverse gradient $\lambda$ was set to $0$, thus we trained the label and domain predictors with ResNet coefficients being influenced only by label predictor. The learning rate was $5\cdot10^{-3}$ for $20$ epochs, $2\cdot10^{-3}$ for the next $40$ epochs and $10^{-3}$ for $40$ more epochs. For the next $300$ epochs the learning rate was kept at $10^{-3}$ and the inverse gradient was gradually increased to $0.1$ in steps $20$ epochs long: $\lambda_n=0.1\cdot(1-1/(1-\exp \rho_n))$ where $\rho_n=-5+\nicefrac{9}{14}\cdot n$, $n=0..14$. For the rest of the training $\lambda$ was set to $0.1$. The training continued for $50$ epochs then the learning rate was dropped to $3\cdot10^{-4}$ for another $50$ epochs and finally the network was allowed to reach the stable state by training for $100$ epochs with learning rate $10^{-5}$.

\section*{Acknowledgements}
The authors are grateful to Yury Melnichek and Valentin Bazarevsky for stimulating discussions, Konstantin Avchaciov, Leonid Menshikov from Gero for numerous discussions and help with the implementation of algorithms. This study was conducted using the UK Biobank Resource, application number 21988.
Funding: This work was funded by Gero LLC.

\section*{Author contributions} TP, EG, BZ and PF designed and performed the numerical modelling, statistical analysis, wrote the manuscript; 

\section*{Competing interests:} 
P.O. Fedichev is a shareholder of Gero LLC.  T.V. Pyrkov, E. Getmantsev, B. Zhurov, and P.O. Fedichev are employees of Gero LLC.  A patent application submitted by Gero LLC on the described methods and tools for evaluating health non-invasively is pending.

\bibliography{ref}

\begin{thebibliography}{10}

\bibitem{hannum}
G.~Hannum, {\it et~al.\/}, {\it Mol. Cell\/} {\bf 49}, 359 (2013).

\bibitem{horvath}
S.~Horvath, {\it Genome Biol.\/} {\bf 14}, R115 (2013).

\bibitem{levine2013bioage}
M.~E. Levine, {\it J. Gerontol. A Biol. Sci. Med. Sci.\/} {\bf 68}, 667 (2013).

\bibitem{zhang2016epigenome}
X.~Zhang, {\it et~al.\/}, {\it Epigenetics\/} {\bf 11}, 750 (2016).

\bibitem{pmid25969563}
S.~Horvath, A.~J. Levine, {\it {J. Infect. Dis.}\/} {\bf 212}, 1563 (2015).

\bibitem{pmid25678027}
S.~Horvath, {\it et~al.\/}, {\it {Aging Cell}\/} {\bf 14}, 491 (2015).

\bibitem{horvath2014obesity}
S.~Horvath, {\it et~al.\/}, {\it Proceedings of the National Academy of
  Sciences\/} {\bf 111}, 15538 (2014).

\bibitem{liu2018phenotypic}
Z.~Liu, {\it et~al.\/}, {\it bioRxiv\/} p. 363291 (2018).

\bibitem{pyrkov2017quantitative}
T.~V. Pyrkov, {\it et~al.\/}, {\it Aging\/} {\bf 10}, 2973 (2018).

\bibitem{pyrkov2018extracting}
T.~V. Pyrkov, {\it et~al.\/}, {\it Scientific reports\/} {\bf 8}, 5210 (2018).

\bibitem{althoff2017large}
T.~Althoff, {\it et~al.\/}, {\it Nature\/} {\bf 547}, 336 (2017).

\bibitem{ganin2016domain}
Y.~Ganin, {\it et~al.\/}, {\it The Journal of Machine Learning Research\/} {\bf
  17}, 2096 (2016).

\bibitem{rajpurkar2017cardiologist}
P.~Rajpurkar, A.~Y. Hannun, M.~Haghpanahi, C.~Bourn, A.~Y. Ng, {\it arXiv
  preprint arXiv:1707.01836\/}  (2017).

\bibitem{ben2010theory}
S.~Ben-David, {\it et~al.\/}, {\it Machine learning\/} {\bf 79}, 151 (2010).

\bibitem{zenin2018identification}
A.~Zenin, {\it et~al.\/}, {\it bioRxiv\/} p. 300889 (2018).

\bibitem{green1983comparison}
M.~S. Green, M.~J. Symons, {\it Journal of chronic diseases\/} {\bf 36}, 715
  (1983).

\bibitem{abbott1985logistic}
R.~D. Abbott, {\it American Journal of Epidemiology\/} {\bf 121}, 465 (1985).

\bibitem{he2016identity}
K.~He, X.~Zhang, S.~Ren, J.~Sun, {\it European conference on computer vision\/}
  (Springer, 2016), pp. 630--645.

\bibitem{tensorflow2015-whitepaper}
M.~Abadi, {\it et~al.\/}, {TensorFlow}: Large-scale machine learning on
  heterogeneous systems (2015). Software available from tensorflow.org.

\bibitem{kingma2014adam}
D.~P. Kingma, J.~Ba, {\it arXiv preprint arXiv:1412.6980\/}  (2014).

\bibitem{gompertz1825nature}
B.~Gompertz, {\it Philosophical transactions of the Royal Society of London\/}
  {\bf 115}, 513 (1825).

\bibitem{makeham1860law}
W.~M. Makeham, {\it The Assurance Magazine and Journal of the Institute of
  Actuaries\/} {\bf 8}, 301 (1860).

\bibitem{st2015risk}
J.~L. St~Sauver, {\it et~al.\/}, {\it BMJ open\/} {\bf 5}, e006413 (2015).

\bibitem{willis2018association}
E.~A. Willis, J.~J. Shearer, C.~E. Matthews, J.~N. Hofmann, {\it PloS one\/}
  {\bf 13}, e0204277 (2018).

\bibitem{jensen2013elevated}
M.~T. Jensen, P.~Suadicani, H.~O. Hein, F.~Gyntelberg, {\it Heart\/} {\bf 99},
  882 (2013).

\bibitem{he2016deep}
K.~He, X.~Zhang, S.~Ren, J.~Sun, {\it Proceedings of the IEEE conference on
  computer vision and pattern recognition\/} (2016), pp. 770--778.

\bibitem{andersen2012health}
S.~L. Andersen, P.~Sebastiani, D.~A. Dworkis, L.~Feldman, T.~T. Perls, {\it
  Journals of Gerontology Series A: Biomedical Sciences and Medical Sciences\/}
  {\bf 67}, 395 (2012).

\bibitem{ioffe2015batch}
S.~Ioffe, C.~Szegedy, {\it arXiv preprint arXiv:1502.03167\/}  (2015).

\bibitem{ioffe2017batch}
S.~Ioffe, {\it Advances in Neural Information Processing Systems\/} (2017), pp.
  1945--1953.

\bibitem{li2018understanding}
X.~Li, S.~Chen, X.~Hu, J.~Yang, {\it arXiv preprint arXiv:1801.05134\/}
  (2018).

\end{thebibliography}
\bibliographystyle{Science}

\end{document}